\documentclass[twocolumn,showpacs,preprintnumbers,amsmath,amssymb]{revtex4}

\usepackage{graphicx}
\usepackage{dcolumn}
\usepackage{bm}

\input epsf

\begin{document}




\title{Magnetization plateaus in the Ising limit of the multiple-spin
exchange model on plaquette chain.}

\author{ V. R. Ohanyan$^{1,2}$ and N. S. Ananikian$^1$ \\[1mm]
{\small \sl $^1$ Department of Theoretical Physics, Yerevan Physics Institute,} \\
{\small \sl Alikhanian Brothers 2, 375036 Yerevan, Armenia}\\[1mm]
{\small \sl $^2$ Chair of Theoretical Physics, Yerevan State University,} \\
{\small \sl Al. Manoogian 1, 375049, Yerevan, Armenia}\\
{\small \sl Russian-Armenian University,}\\
{\small \sl 123 Hovsep Emin St, Yerevan, 375051, Armenia}
 \\[1mm]}

\begin{abstract}
 We consider the Ising spin system, which stems out from the
 corresponding Multiple-spin exchange (MSE) Hamiltonian, on the special
 one--dimensional lattice, diamond-plaquette chain. Using the technique
 e of transfer-matrix we obtain the exact expression for system free energy
 with the aid of which we obtain the magnetization function. Analyzing
 magnetization curves for varies values of temperature and couplin
 constants we found the magnetization plateaux at $1/3$ and $2/3$ of the full moment.
 The corresponding microscopic spin configurations are unknown by virtue of
 high frustration.
 \end{abstract}
\maketitle

\baselineskip=13.07pt
 The Heisenberg model is widely recognized as a lattice model for
 magnetism of materials. However, this model by no means is
 universal, because it based on several assumptions. One of these
 assumptions consists in the pair character of exchange
 interactions. This means that only the exchange processes of no
 more than two particles (nearest-neighbour or next-nearest-neighbour and so
 on) are taken into consideration:
 \begin{eqnarray}
 {\mathcal{H}}_{Heis}=2J \sum_{\langle i, j \rangle} P_{ij}.
 \label{Heis}
 \end{eqnarray}
 Here $P_{ij}$ are the pair exchange operators, which implement
 the transposition of two spin states in $i$-th and $j$-th sites
 of the lattice:
 \begin{eqnarray}
 P_{ij}|\xi_i \rangle \otimes | \xi_j \rangle = | \xi_j \rangle \otimes |\xi_i
 \rangle. \label{perm}
 \end{eqnarray}
For the SU(2) spins and $s=1/2$ the expression for $P_{ij}$ is
\begin{equation}
 P_{ij} = \frac 12\left( 1+\mbox{\boldmath $\sigma$}_i \cdot
\mbox{\boldmath $\sigma$}_j\right) , \label{P2}
 \end{equation}
 $\mbox{\boldmath $\sigma$}_i$ are the Pauli matrices.
 The generalization of this picture is known since 60-s
 \cite{Thou}and called the multiple-spin exchange (MSE) model.
 This model describes the magnetism of the system of almost
 localized fermions via the concept of many particle permutation.
 The general form of MSE Hamiltonian is
 \begin{equation}
{\mathcal{H}}_{ex}=-\sum_{n,\alpha }J_{n\alpha }\left( -1\right)
^p P_n . \label{MSE_H}
\end{equation}
Here $P_n$ denotes the $n-$particle cyclic permutation operator,
$J_n$ is corresponding exchange energy and $p$ is the parity of
the permutation, which indicates how much pair transpositions
contains in the given cyclic permutation. Many peculiar magnetic
and thermodynamical properties of $^3$He adsorbed on graphite
surface can be understood only within the framework of MSE model
\cite{He}. Recently the significant role of MSE interaction was
revealed in low-dimensional cuprate compounds \cite{Cu}, what initiated the
interest toward the two-leg spin ladders with four-spin cyclic interaction
\cite{Lad}. The MSE model by itself exhibits rich phase structure even
at classical level \cite{MSE}. Another interesting feature of MSE model is the possibility of complex magnetic behavior,
including such a phenomena as magnetization plateaux, which was established in the
 MSE model on triangular lattice \cite{Mom}. The study of the magnetization plateaux
 \cite{Pla} is the one of the main directions of nowaday investigations on macroscopic non--trivial
 quantum effects in condensed matter physics, which have a number of fundamental
 and applied importance. Despite the purely quantum origin of this effect it was shown recently
 that, magnetization plateaux can appear in the Ising spin systems also \cite{Str,Oha,Ara}, exhibiting in some cases fully qualitative correspondence with its Heisenbegr
 counterpart \cite{Oha}. The latter fact is very important because it can serve
 for more profound understanding of magnetization plateaux physics, can provide it with new methods
 and can initiate a search of novel magnetic materials with huge axial anisotropy.
\begin{figure}
  \includegraphics[width=9cm]{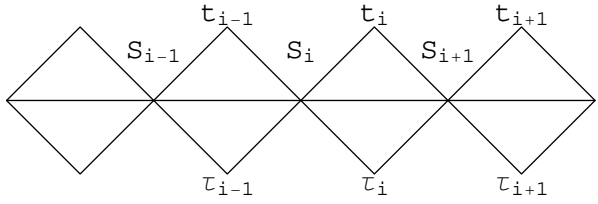}
  \caption{\label{fig1} The diamond
 plaquette spin chain. }
         \end{figure}

\section{The Model}
 We consider a one-dimensional lattice consists of corner shared so--called
 diamond plaquettes (See Fig. ~\ref{fig1}). The diamond plaquette is a square plaquette
 of 4 spins with nearest neighbor interaction and additional bound, connecting two
 opposite spins. If we consider the Hamiltonian of MSE model and restrict ourself
 with the two-, three-, and four--spin exchanges, which is the
 simplest case, we arrive at he following Hamiltonian:

\begin{widetext}
\begin{eqnarray}
{\mathcal{H}}&=&J \sum_{\left(i,j \right)}\mbox{\boldmath
$\sigma$}_i \cdot \mbox{\boldmath $\sigma$}_j+K \sum_p h_p-\mu H
\sum_i {\sigma _i}^z, \\ \nonumber
 h_p &=& \sum_{1\leq i \leq j
\leq 4}\mbox{\boldmath $\sigma$}_i \cdot \mbox{\boldmath
$\sigma$}_j+\left(\mbox{\boldmath $\sigma$}_1 \cdot
\mbox{\boldmath $\sigma$}_2\right)\left(\mbox{\boldmath
$\sigma$}_3 \cdot \mbox{\boldmath $\sigma$}_4\right)+
 \left(\mbox{\boldmath
$\sigma$}_1 \cdot \mbox{\boldmath
$\sigma$}_4\right)\left(\mbox{\boldmath $\sigma$}_2 \cdot
\mbox{\boldmath $\sigma$}_3\right)-\left(\mbox{\boldmath
$\sigma$}_1 \cdot \mbox{\boldmath
$\sigma$}_3\right)\left(\mbox{\boldmath $\sigma$}_2 \cdot
\mbox{\boldmath $\sigma$}_4\right), \label{Ham:4}
\end{eqnarray}
\end{widetext}
where $J=J_3-J_2/2$ and $K=-J_4/4$. This model have been
successfully applying for describing the properties of solid
$^3$He films. In our case we omit the off--diagonal part of the
Hamiltonian \ref{Ham:4} what is equivalent to the formal
replacement of all spin operators by simple Ising variables which
take values $\pm 1$. Putting the system, describing by the
Hamiltonian \ref{Ham:4} onto the lattice, depicted in Figure
\ref{Fig1}, we get the following model:
\begin{widetext}
\begin{eqnarray}
-\beta{\mathcal{H}}&=& \sum_i \alpha_1 s_i s_{i+1}+\alpha_2
\nonumber \left(s_i+s_{i+1} \right) \left(t_i+\tau_i\right)+
\alpha_3\left(t_i \tau_i+s_i s_{i+1} t_i
\tau_i\right)+h\left(s_i+t_i+\tau_i\right), \\  \label{ham:f}
\end{eqnarray}
\end{widetext}
where
\begin{eqnarray}
 &&\alpha_1=\beta\left(J_3-J_2/2-J_4/4\right), \\  \nonumber
&&\alpha_2=\beta\left(J_3/2-J_2/2-J_4/4\right), \\  \nonumber
&&\alpha_3=-\beta J_4/4.
\end{eqnarray}
Here the Ising spins placed in the corners of diamond plaquette
chain denoted by $t_i$ and $\tau_i$ and spins placed in the middle
line by $s_i$. The system allows the exact calculation of the
partition function, which can be represented a sa trace of the
N-th power of the corresponding transfer-matrix \cite{Bax}

\begin{eqnarray}
Z=\sum_{\left(s,t,\tau \right)}e^{-\beta{\mathcal{H}}}=\mbox{Sp}
T^N.
\end{eqnarray}

The transfer--matrix is
\begin{widetext}
\begin{center}
\begin{eqnarray}
\mathbf{T}= \left ( \begin{array}{cc}
      2e^{\alpha_1 +h}\left(e^{2\alpha_3}\cosh(4\alpha_2+2h)+e^{2\alpha_3}\right), &
      2e^{-\alpha_1}\left(\cosh(2h)+1 \right)\\
      \\
      2e^{-\alpha_1}\left(\cosh(2h)+1 \right), &
      2e^{\alpha_1 -h}\left(e^{2\alpha_3}\cosh(4\alpha_2-2h)+e^{2\alpha_3}\right)\label{TM}
      \end{array}
\right).
\end{eqnarray}
\end{center}
\end{widetext}
As usually to calculate partition function we should obtain the
eigenvalues of the transfer matrix, more precisely we need only
the maximal eigenvalue, because in the thermodynamical limit ($ N
\to \infty $) only this one survives. So, $Z_N=\lambda_{max}^N$
The maximal eigenvalue is

\begin{widetext}
\begin{eqnarray}
\lambda=A\cosh(h)+B\cosh(3h)+\sqrt{\sum_{k=o}^3 C_k \cosh(2kh)},
\label{l}
\end{eqnarray}
\end{widetext}
where the coefficients are
\begin{widetext}
\begin{eqnarray}
&&A=e^{\alpha_1-4\alpha_2+2\alpha_3}+2e^{\alpha_1-2\alpha_3},\\
&&B=e^{\alpha_1+4\alpha_2+2\alpha_3}, \\
&&C_0=e^{2\alpha_1}\left(6e^{-4\alpha_1}+2e^{-4\alpha_2}-2e^{-4\alpha_3}
-e^{4\alpha_3} \cosh(8\alpha_2)\right), \\
&&C_1=e^{2\alpha_1+4\alpha_3} \left(1+\frac{1}{2}
e^{-8\alpha_2}+2e^{-8\alpha_3}\right)-4e^{2\alpha1}\cosh(4\alpha_2)
+8e^{-2\alpha_1}, \\
&&C_2=e^{2\alpha_2}\left(2e^{-4\alpha_1}+2e^{4\alpha_2}-e^{4\alpha_3}\right), \\
&&C_3=e^{2\alpha_1+8\alpha_2+4\alpha_3}.
\end{eqnarray}
\end{widetext}
Having the partition function we can obtain the free energy of the
system per one spin in the thermodynamical limit:

\begin{eqnarray}
f=-\frac{1}{\beta}\lim_{N \rightarrow \infty }\frac{\log
\lambda^N}{3N}= -\frac{1}{3\beta}\log \lambda .
\end{eqnarray}
Then , using the conventional thermodynamical relations
$m=-\left(\frac{\partial f}{\partial H}\right)$, we obtain the
magnetization per spin as s explicit function
\begin{widetext}
\begin{eqnarray}
m=\frac{A\sinh(h)+3B\sinh(3h)+\frac{\sqrt{\sum_{k=1}^3 k C_k
\sinh(2kh)}}{\sqrt{\sum_{k=o}^3 C_k \cosh(2kh)}}}
{3\left(A\cosh(h)+B\cosh(3h)+\sqrt{\sum_{k=o}^3 C_k \cosh(2kh)}
\right)}.
\end{eqnarray}
\end{widetext}
\begin{figure}
 \begin{center}
  \includegraphics{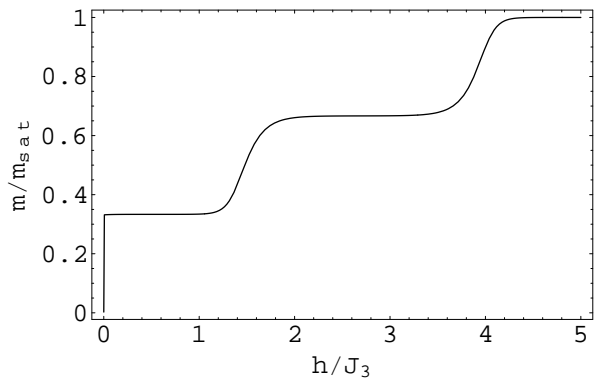}
  \caption{\label{fig2} The
magnetization curve for Ising plaquette chain at $T/J_3=0.17$,
$J_2/J_3=1.5$ and $J_4/J_3=1.7$. }
           \end{center}
         \end{figure}


Having this function we can draw the plots of the magnetization
processes for all finite temperatures and arbitrary values of
coupling constants. First of all, due to its geometry the system
is highly frustrated in case of antiferromagnetic effective
coupling constants. This means that even at $T=0$ the ground state
of the system is disordered, because the arrangement of spins on
lattice precludes satisfying all interaction simultaneously.
However, it is possible to choose such MSE coupling constants at
which the frustration will be partially or entirely removed in the
Ising limit. Among the variety of magnetization curves obtained
for different sets of $J_2$, $J_3$ and $J_4$ the most remarkable
is the one with two magnetization plateaux at $m=1/3$ and $m=2/3$
in the units of full moment(Fig. \ref{fig2}). At that region of
coupling constants the system is highly frustrated and it is not
so easy to determinate the microscopic spins configurations
corresponding to these plateaux. Apparently they are some complex
periodic structures with spatial period at lest $3$ for
$1/3$-plateau and $6$ for $2/3$-plateau. The investigation of the
entire MSE model on diamond plaquette chain
 may change the picture obtained by us drastically. On the one
 hand, the appearance of other plateaux is possible, on the other hand,
 the plateaux pertinent the Ising system might not survive in quantum case.
 The typical example is the simple $S=1/2$ Heisenberg chain which
 is gapless, whereas the corresponding Ising system exhibits the
 plateau at $m=0$. Let us mention also that apparently among the
 overdoped RCuO$_2+x$ (R=Y, La, {\em etc}.) compounds \cite{Cav}
 the ones are possible whose magnetic lattice is analogous to that
 considered here.



\begin{thebibliography}{99}

\bibitem{Thou} D.J. Thouless, Proc. Phys. Soc.(London) \textbf{86},893 (1965).

\bibitem{He} G. Misguich, B. Bernu, C. Lhuillier, and C. Waldtmann,
Phys. Rev. Lett. \textbf{81}, 1098 (1998);
 G. Misguich, C. Lhuillier, B. Bernu, and C. Waldtmann, Phys. Rev. \textbf{B} 60, 1064 (1999);
 W. LiMing, G. Misguich, P.
Sindzingre, and C. Lhuillier, Phys. Rev. \textbf{B} 63, 6372
(2001).

\bibitem{Cu} R. Coldea {\em et al.}, Phys. Rev. Lett. \textbf{86}, 5377 (2001);
M. Matsuda, K. Katsumata, R. S. Eccleston, S. Brehmer, and H.-J.
Mikeska, Phys. Rev. \textbf{B} 62, 8903 (2000); M. Windt {\em et
al.}, Phys. Rev. Lett. \textbf{87}, 127002 (2001); K. P. Schmidt,
C. Knetter, and G. S. Uhrig, Europhys. Lett. \textbf{56}, 877
(2001).

\bibitem{Lad} M. Muller, T. Vekua and H.-J. Mikeska,Phys. Rev. \textbf{B} 66, 134423 (2002);
 V. Gritsev, B. Normand, and D.
Baeriswyl, Phys. Rev. \textbf{B} 69, 094431 (2004) and references
therein.
\bibitem{MSE} K. Kubo and T. Momoi,  Z. Phys. \textbf{B} 103, 485 (1997).
\bibitem{Mom} T. Momoi, H. Sakamoto, and K. Kubo, Phys. Rev. \textbf{B} 59, 9491 (1999).
\bibitem{Pla} D. C. Cabra, M. D. Grynberg, A. Honecker, and P. Pujol
in {\em Condensed Matter Theories}, vol. 16. eds. S. Hernandez and
J. W. Clark, p.17 (Nova Science Publishers, New York, 2000); A.
Honecker, J. Schulenburg and J. Richter, J. Phys.: Condens. Matter
\textbf{16}, S749 (2004) and references therein.
\bibitem{Str} J. Strecka and M. Jascur, J. Phys.: Condens. Matter \textbf{15},4519 (2003).
\bibitem{Oha} V. R. Ohanyan and N. S. Ananikian, Phys. Lett. \textbf{A} 307, 76 (2003).
\bibitem{Ara} T. R. Arakelyan, V. R. Ohanyan, L. N. Ananikyan, N. S. Ananikian, and
M. Roger, Phys. Rev. \textbf{B} 67, 024424 (2003).
\bibitem{Bax} R. Baxter, {\em Exactly Solved Models in Statistical Mechanics}
(Academic Press, 1982).
\bibitem{Cav} R. J. Cava, {\em et al}., J. Solid State Chem.\textbf{104}, 437 (1993).




\end{thebibliography}
\end{document}